\documentclass[11pt,a4paper]{article}
\usepackage{arxiv}
\usepackage[T1]{fontenc}    %
\usepackage{hyperref}       %
\usepackage{url}            %
\usepackage{booktabs}       %
\usepackage{amsfonts}       %
\usepackage{nicefrac}       %
\usepackage{microtype}      %
\usepackage{lipsum}		%
\usepackage{graphicx}
\usepackage{doi}

\usepackage{amsfonts}

\usepackage[font=small,labelfont=bf]{caption}

\usepackage{afterpage}
\usepackage{algorithmic}
\usepackage{array} %

\usepackage{booktabs} %
\usepackage{color}
\usepackage{graphbox}
\usepackage{graphicx}

\usepackage{mathtools}
\usepackage{multirow}
\usepackage{url}
\usepackage{tabularx}
\usepackage{textcomp}
\usepackage{tikz}
\usepackage{pifont} %
\usepackage{colortbl}
\usepackage{flushend}
\usepackage[normalem]{ulem}
\useunder{\uline}{\ul}{}
\usepackage{cleveref} 
\usepackage[T1]{fontenc}

\usepackage{adjustbox}
\usepackage{array}
\usepackage[strings]{underscore}

\newcommand*\rot{\rotatebox{90}}
\newcommand*\OK{\ding{51}}

\newcommand{\method}{configuration method}

\newcommand{\confsys}{configuration system}

\newcommand{\suc}{SUC}
\newcommand{\policy}{policy}
\newcommand{\goal}{goal}

\newcommand{\systemquality}{system quality}
\newcommand{\systemqualities}{system qualities}
\newcommand{\systemqualitymodel}{system quality model}
\newcommand{\observation}{observation}

\newcommand{\sample}{sample}

\newcommand{\methodqualities}{configuration method qualities}
\newcommand{\catdimension}{quality dimension}

\newcommand{\run}{run}

\newcommand{\cc}{\textbf{C}}
\newcommand{\ca}{\textbf{A}}
\newcommand{\ct}{\textbf{T}}

\newcommand{\manitestbeds}{\textbf{T3}}

\newcommand{\fix}[1]{}%
\newcommand{\tactic}[1]{\paragraph{\textbf{#1}}}
\newcommand{\cfix}[1]{}%
\newcommand{\head}[1]{\noindent\textbf{#1}}
\definecolor{darkblue}{rgb}{0.122, 0.435, 0.698}
\definecolor{grey}{gray}{0.30}
\definecolor{lightblue}{rgb}{0, 0.8, 0.99}

\definecolor{sky1a}{HTML}{cfe0e8}
\definecolor{sky1b}{HTML}{b7d7e8}
\definecolor{sky1c}{HTML}{87bdd8}
\definecolor{sky1d}{HTML}{daebe8}

\definecolor{sky2a}{HTML}{bccad6}
\definecolor{sky2b}{HTML}{8d9db6}
\definecolor{sky2c}{HTML}{667292}
\definecolor{sky2d}{HTML}{f1e3dd}

\definecolor{living_coral}{HTML}{FF6F61}
\definecolor{spearmint}{HTML}{64BFA4}

\definecolor{teal}{HTML}{00A08F}
\definecolor{marine}{HTML}{124653}
\definecolor{yellow}{HTML}{FEE074}
\definecolor{orange}{HTML}{FF9469}
\definecolor{pink}{HTML}{FE8d8F}

\begin{document}

\title{Synthesizing Configuration Tactics for Exercising Hidden Options in Serverless Systems}
\author{Jörn Kuhlenkamp\\
ISE, TU Berlin, Germany\\
\texttt{jk@ise.tu-berlin.de} \and
\underline{Sebastian Werner}\footnote{Corosponding Author}\\
ISE, TU Berlin, Germany\\
\texttt{sw@ise.tu-berlin.de} \and
Chin Hong Tran\\
ISE, TU Berlin, Germany\\
\texttt{ct@ise.tu-berlin.de} \and
Stefan Tai \\
ISE, TU Berlin, Germany\\
\texttt{st@ise.tu-berlin.de}}

\date{}

\renewcommand{\undertitle}{preprint of \url{https://doi.org/10.1007/978-3-031-07481-3_5}}
\renewcommand{\shorttitle}{\textit{arXiv} Template}

\hypersetup{
pdftitle={Synthesizing Configuration Tactics for Exercising Hidden Options in Serverless Systems},
pdfsubject={q-bio.NC, q-bio.QM},
pdfauthor={J.~Kuhlenkamp,S.~Werner,C.~Hong,S.~Tai},
pdfkeywords={preprint,serverless system \and configuration methods  \and design tactics},
}

\maketitle
\cfix{\vspace{-1em}}
\thispagestyle{plain}
\pagestyle{plain}
\begin{abstract}
A proper configuration of an information system can ensure accuracy and efficiency, among other system objectives. 
Conversely, a poor configuration can have a significant negative impact on the system's performance, reliability, and cost. 
Serverless systems, which are comprised of many functions and managed services, especially risk exposure to misconfigurations, with many provider- and platform-specific, often intransparent and ‘hidden’ settings.
In this paper, we argue to pay close attention to the configuration of serverless systems to exercise options with known accuracy, cost and time. Based on a literature study and long-term serverless systems development experience, we present nine tactics to unlock potentially neglected and unknown options in serverless systems. 

\end{abstract}

\section{Introduction}\label{sec:introduction}
Serverless computing is a new cloud provisioning model appealing to cloud consumers due to increased automation of operational tasks, instant scalability with incoming workload, and no costs for idle computing resources. %
Consequently, serverless computing promises developers the means for delivering more features for applications in shorter lead times at a lower cost~\cite{kuhlenkamp-2020-ic2e-ifs_and_buts}.

In the serverless programming model~\cite{2019-Jonas-arxiv-BerkeleyViewOnServerlessComputing}, developers build serverless systems as event-driven compositions of (cloud) functions and additional, managed cloud services, such as object storage, messaging queues, and databases.
Cloud functions run in a Function-as-a-Service platform such as AWS Lambda, Google Cloud Functions, or Apache OpenWhisk.
However, academia and industry report that client-side quality in terms of performance, reliability, and execution costs frequently and significantly mismatches developers' expectations due to poor configurations~\cite{kuhlenkamp-2019-ucc-opstasks,2020-Kuhlenkamp-ACR-All_But_One} making ignoring configuration particularly risky.

Ironically, configuring serverless systems "accurately" requires developers to understand a cloud platform's quality-sensitive configuration options hidden by the programming model's high abstraction.
To this end, related work proposes configuration methods \cite{2018-Saha-CLOUD-EMARS,2018-Tarek-SEC-Costless,2020-Akhtar-Infocom-COSE,2020-Eismann-ICPE-Workflow_Costs}. However, these configuration methods may not live up to developers' expectations, as they focus on accuracy but tend to neglect required cost and time. Furthermore, 
accuracy with these methods typically depends on implicit assumptions regarding a system's architecture and deployment environment.
We find it not convincing to assume that developers that select the serverless computing paradigm to leverage cost and lead-time benefits buy into a configuration approach that demands unknown cost and time investments.

In this context, we define the research question: \textbf{How can developers configure serverless systems with predictable accuracy, cost, and time?}
Towards this, we present two contributions, (i) a literature review that indicates that accuracy, cost and time of (current) \method{}s are too in-transparent, and (ii) nine tactics for designing future \method{}s  with known accuracy, cost and time effectively synthesizing the state-of-the-art.

\section{Configuration Fundamentals\label{sec:cat}}
Through introducing fundamentals of system \method{}s', we lay the foundation for reviewing existing related work (Sec. \ref{sec:rw}), introducing \textbf{tasks}, and \textbf{quality dimensions}, see also Figure~\ref{fig:overview}.

\hspace*{-0.5cm}
\begin{minipage}{\textwidth}
  \begin{minipage}[b]{0.65\textwidth}
    \includegraphics[width=\textwidth]{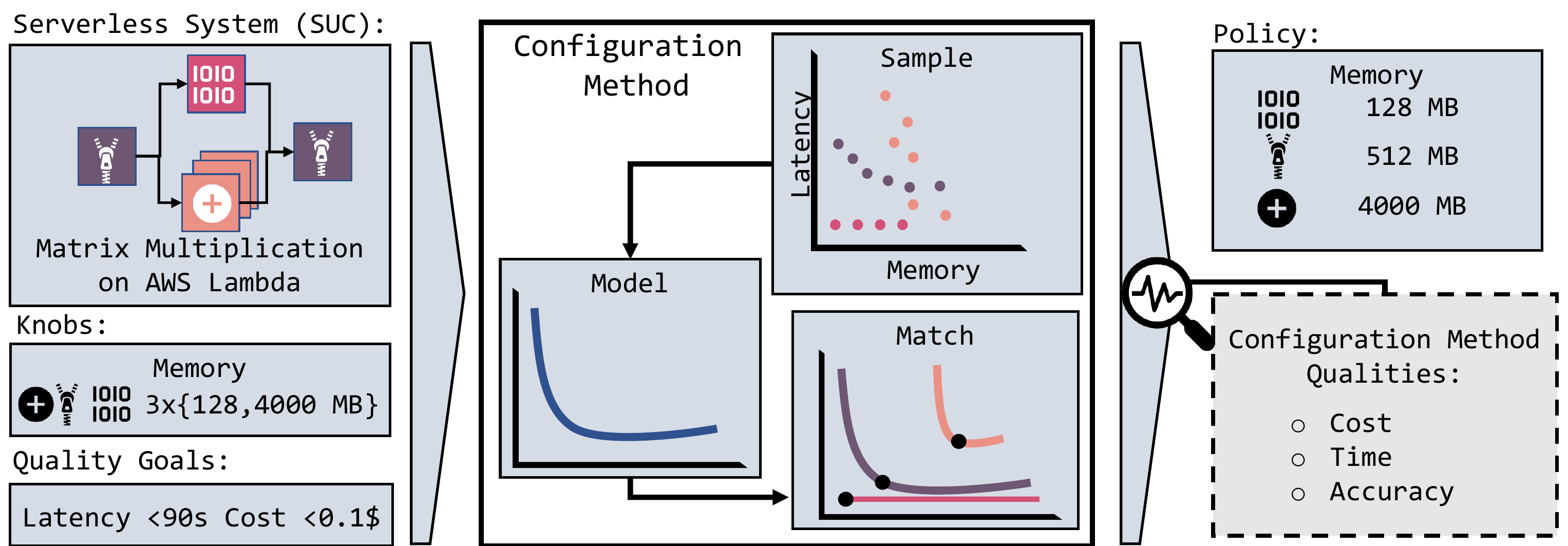}
    \captionof{figure}{Exemplary \method{} on Lambda.}
    \label{fig:overview}
  \end{minipage}
  \begin{minipage}[b]{0.34\textwidth}
    \resizebox{\textwidth}{!}{%
      
\newcolumntype{R}[2]{%
    >{\adjustbox{angle=#1,lap=\width-(#2)}\bgroup}%
    l%
    <{\egroup}%
}
\renewcommand\rot{\multicolumn{1}{R{45}{1em}}}

\begin{tabular}{@{} cl*{5}c | *{2}c | *{3}c @{}}
        & & \rot{Memory} &\rot{Batch} & \rot{Fusion} & \rot{Placement} & \rot{Concurrency} & \rot{Bounds} & \rot{Weights} & \rot{Isolated} & \rot{Chain} & \rot{Complex} \\
        \cmidrule{2-12}
        & Saha\cite{2018-Saha-CLOUD-EMARS}                  & \OK &   - &   - &   - &   - &   - &   - & \OK &   - &     - \\
        & Christoforou\cite{2018-Christoforou-ESSCA-Res_Mgnt}         & \OK & \OK &   - &   - & \OK &   - &   - & \OK &   - &     - \\
        & Ali\cite{2020-Ali-SC-Batch}                        &   - & \OK &   - &   - &   - & \OK &   - & \OK &   - &     - \\
        & Elgamal\cite{2018-Tarek-SEC-Costless}                & \OK &   - & \OK & \OK &   - & \OK &   - & \OK & \OK &   \OK \\
        & Schuler\cite{2020-Schuler-arxiv-Concurrency_Agent}                          &   - &   - &   - &   - & \OK &   - &   - & \OK &   - &     - \\
        & Akhtar\cite{2020-Akhtar-Infocom-COSE}              & \OK &   - &   - & \OK &   - & \OK &   - & \OK & \OK &     - \\
        & Tariq\cite{2020-Tariq-SAC-Sequoia}                 &  -  &   - &   - &  -  & \OK & \OK &   - & \OK & \OK & \OK \\
        & S\'anchez-A.\cite{2020-Sanchez-Middleware-Primula}  &  -  &   - &   - &  -  & \OK & \OK  &   - &  -  &  -  &  \OK \\    
        & Eismann\cite{2021-Eismann-Preprint-Sizeless}       & \OK &   - &   - &   - &   - &   - & \OK & \OK &   - &     - \\
        & Sedefouglu\cite{2021-Sedefouglu-SAC-Sizing}        & \OK &   - &   - &   - &   - &   - &   - & \OK &   - &     - \\
        \cmidrule{2-12}
        & First-Author & \multicolumn{5}{c}{Knob} & \multicolumn{2}{c}{Goal} & \multicolumn{3}{c}{Composition} \\[1ex]
\end{tabular}

      }
      \captionof{table}{Comparison of \method{}s.}
      \label{tab:rw}
    \end{minipage}
 \end{minipage}

\subsection{Tasks\label{sec:cat:method}}
Developers configure systems to meet quality goals.
To that extent, a \method{} takes as input a \textit{system under configuration} (\suc{}), knobs, e.g., a configuration parameter, and \goal{}s, e.g., a target throughput, and outputs a \policy{}, e.g., a set of configuration parameters that should meet desired \goal{}s.
In Figure~\ref{fig:overview}, we show a running example of sizing "memory" for three functions implementing matrix multiplication on AWS Lambda, with two \goal{}s: client-side request-response latency below 90s and marginal execution cost below 0.1 USD. 

To meet both \goal{}s, a \method{}s needs to find values for \textit{knob}s, e.g., the memory size of the three functions in our example.
A \suc{} can expose knobs on the platform- or application-level.
However, application-level knobs often require developers to modify code, e.g., changing libraries or the function composition.
With increasing maturity, a platform typically assimilates relevant application-level knobs to expose them as a feature to foster accessibility~\cite{2021-Werner-ICSOC-CoDesign}.

A \textit{\goal{}} is an expression of a developer's preferences for a \systemquality{}, e.g., a target latency range.
Note that \method{}s assume that a \suc{} allows observing \systemqualities{} included in the \goal{}s.
A \textbf{\method{}} comprises three high-level tasks: \textit{match}, \textit{model}, and \textit{sample} resulting in a \textit{\policy{}} that fits a \textit{\goal{}}. 
\head{Matching} finds a \policy{} that fits provided \goal{}s accurately.
Thus, matching requires searching all possible configurations.
To assess how accurately a \policy{} matches \goal{}s, matching uses a \textit{\systemqualitymodel{}} that captures the cause-effect relationship between a \policy{} and \goal{}s, e.g., cost and latency as a function of memory size.
\head{Modelling} is the task of creating such a \systemqualitymodel{}.
Computing a realistic \systemqualitymodel{} typically requires collecting \sample{}s under realistic assumptions.
\head{Sampling} is the task of obtaining such \sample{}s.
A \sample{} is a data record including a \policy{} and measured \systemquality{}, e.g., 8s latency and 0.2 USD cost for 1024MB memory size.
While it is possible to create \sample{}s using different approaches, reliable results require experimentation, i.e., making \observation{}s on a \suc{} in a controlled environment.
\vspace{-1em}
\subsection{Quality Dimensions\label{sec:cat:dimensions}}
Quality dimensions quantify a \method{}'s quality, and are hence not to confuse with system qualities used in specifying \goal{}s.
In order to compare and evaluate \method{}s, we present the three most relevant \methodqualities{} in detail.

\head{Accuracy (\ca{})} quantifies how closely a \method{}'s computed \policy{}s fit \goal{}s.
Goals including multiple \systemqualities{} typically require an aggregation function such as a weighted distances function for comparison.
Another aspect is stability, which describes the ability of a \method{} to generalize, i.e., maintaining accuracy for changing \suc{}s.
\head{Costs (\cc{})} comprises all monetary costs for executing a \method{}.
It is vital to evaluate not only matching cost but also the modelling and sampling cost.
Otherwise, a \method{} implicitly claims that \systemqualitymodel{}s are reusable and generalize well over arbitrary \suc{}s.
\head{Time (\ct{})} is the difference between applying a computed \policy{} and starting a \method{}.
A \method{} that is fully automated through a \confsys{} typically allows to easily quantify time for configuration requests.
Time for all tasks is relevant because a \method{} can cache samples and \systemqualitymodel{}s.
\cfix{\vspace{-1em}}

\section{Literature Review}\label{sec:rw}
\cfix{\vspace{-0.5em}}
Building on Sec.~\ref{sec:cat}, we analyse existing proposals for configuring serverless systems in a structured literature review (SLR) \cite{kitchenham-2007-slr_guidelines}, extending the scope of serverless computing's literature reviews on other topics~\cite{2018-Kuhlenkamp-WoSC-Survey,2018-Leitner-JoSS-FaasInPractice,2020-Taibi-CLOSER-Function_Patterns,2020-Scheuner-JSS-FaaS_Perf_Study}.
We seed our search for publications in existing serverless computing literature datasets~\cite{2020-Scheuner-JSS-FaaS_Perf_Study,2018-Kuhlenkamp-WoSC-Survey} including only scientific publications from the years 2018-2021.
Each publication includes at least one configuration method in the context of serverless computing.
Table~\ref{tab:rw} summarizes the final set of publications~\cite{2018-Saha-CLOUD-EMARS,2018-Christoforou-ESSCA-Res_Mgnt,2020-Ali-SC-Batch,2018-Tarek-SEC-Costless,2020-Sanchez-Middleware-Primula,2020-Schuler-arxiv-Concurrency_Agent,2020-Akhtar-Infocom-COSE,2021-Eismann-Preprint-Sizeless} in chronological order of appearance.

\subsection{Results}

\head{SUC}
A single client's event typically triggers the execution of multiple downstream functions that form a function composition.
We differentiate between approaches for \textit{isolated} functions \cite{2018-Saha-CLOUD-EMARS,2020-Ali-SC-Batch,2021-Eismann-Preprint-Sizeless,2021-Sedefouglu-SAC-Sizing}, function \textit{chains} \cite{2018-Tarek-SEC-Costless,2020-Akhtar-Infocom-COSE}, and \textit{complex} compositions \cite{2018-Christoforou-ESSCA-Res_Mgnt}.
While a function chain executes multiple functions sequentially, a complex composition includes switching and parallel executions.
Approaches that configure functions in isolation and neglect the composition logic are inclined to propose solutions that do not closely match goals.
Complex compositions include perfectly parallelizable execution flows.%
Some proposals are tightly coupled to a specific system,e.g., \cite{2020-Sanchez-Middleware-Primula} is tightly coupled to the concrete implementation of serverless sorting and \cite{2020-Schuler-arxiv-Concurrency_Agent} to the KNative platform~\cite{2019-Kaviani-WoSC-Knative} hurting general applicability of approaches.
First standalone configuration systems begin to emerge, i.e., Tariq et al.~\cite{2020-Tariq-SAC-Sequoia} provide a middleware called Sequoia for their matching algorithm.
Such approaches seems promising because they make fewer assumptions on a \suc{}s software and deployment architecture. 

\vspace{1em}
\head{Knobs}
Related work addresses five classes of knobs: memory, batch, fusion, placement, and concurrency.
The \textit{memory} knob universally represents setting computing resources for a function (sizing) \cite{2018-Saha-CLOUD-EMARS,2018-Christoforou-ESSCA-Res_Mgnt,2018-Tarek-SEC-Costless,2020-Akhtar-Infocom-COSE,2021-Eismann-Preprint-Sizeless,2021-Sedefouglu-SAC-Sizing}.
Some approaches use small domain size \cite{2021-Eismann-Preprint-Sizeless,2021-Sedefouglu-SAC-Sizing} hiding platform options.
The \textit{batch} knob comes in the two variants event baching and input batching.
\textit{Event batching} \cite{2020-Ali-SC-Batch} collects multiple events before sending them to a single slot, i.e., VM or container.
\textit{Input batching} \cite{2018-Christoforou-ESSCA-Res_Mgnt} collects various inputs in a single event before execution in a single slot.
The \textit{fusion} knob \cite{2018-Tarek-SEC-Costless} merges a set of function handlers into a single one.
In contrast, the \textit{placement} \cite{2018-Tarek-SEC-Costless,2020-Akhtar-Infocom-COSE} knob sets the target deployment environment of a function, e.g., a data center and edge location.
Finally, the \textit{concurrency} knob determines the number of events that a function executes in parallel, including two variants.
\textit{Function concurrency} \cite{2018-Christoforou-ESSCA-Res_Mgnt,2020-Tariq-SAC-Sequoia} determines the sum of events that all slots can process at the same time.
\textit{Slot concurrency} \cite{2020-Schuler-arxiv-Concurrency_Agent} determines the number of events that a single slot can process concurrently, i.e., similarly to multi-threading.

\vspace{1em}
\head{Goals}
From a developer's perspective, all tuning knobs can influence multiple system qualities, including execution latency, cost, and reliability, but existing proposals commonly focus on a subset of these qualities including throughput \cite{2020-Schuler-arxiv-Concurrency_Agent}, execution latency \cite{2018-Saha-CLOUD-EMARS,2018-Christoforou-ESSCA-Res_Mgnt,2020-Ali-SC-Batch,2018-Tarek-SEC-Costless,2020-Sanchez-Middleware-Primula,2020-Akhtar-Infocom-COSE,2021-Eismann-Preprint-Sizeless,2021-Sedefouglu-SAC-Sizing}, execution cost \cite{2018-Christoforou-ESSCA-Res_Mgnt,2020-Ali-SC-Batch,2018-Tarek-SEC-Costless,2020-Akhtar-Infocom-COSE,2021-Eismann-Preprint-Sizeless,2021-Sedefouglu-SAC-Sizing}, and reliability \cite{2020-Tariq-SAC-Sequoia}.
Absolute quality bounds \cite{2018-Tarek-SEC-Costless,2020-Akhtar-Infocom-COSE,2021-Sedefouglu-SAC-Sizing} and relative quality weights \cite{2021-Eismann-Preprint-Sizeless} are both usefully means for expressing developer's quality preferences that are not considered in combination.
Hence, we argue that approaches (currently) limit developers' ability to express quality preferences holistically.

\vspace{1em}
\head{Quality Dimensions}
Not all proposals explicitly discuss quality dimensions \cite{2018-Saha-CLOUD-EMARS}.
Multiple proposals evaluate accuracy without cost and time \cite{2020-Ali-SC-Batch,2018-Christoforou-ESSCA-Res_Mgnt,2018-Tarek-SEC-Costless,2020-Sanchez-Middleware-Primula,2020-Schuler-arxiv-Concurrency_Agent}.
For example, \cite{2018-Christoforou-ESSCA-Res_Mgnt} report high accuracy by quantitative comparisons with an exhaustive search.
We argue that extensive sampling and profiling implies high cost \cite{2020-Ali-SC-Batch}, and potentially time \cite{2018-Christoforou-ESSCA-Res_Mgnt}.
Schuler et al.~\cite{2020-Schuler-arxiv-Concurrency_Agent} use reinforcement learning that requires 150-600 iterations to stabilize in a policy.
For sampling, each iteration executes 500 requests in flight for 30 seconds.
Hence, we infer a stabilization time between 75min and 5h and cost of stabilization time multiplied by average throughput.
Tarek et al. \cite{2018-Tarek-SEC-Costless} propose the knob function fusion that implies changing the application logic of a \suc{}, which makes estimating cost and time of the approach difficult.
Most proposals do not discuss how they handle changes to a \suc{}.
Akhtar et al. \cite{2020-Akhtar-Infocom-COSE} generate separate quality models for individual functions using Bayesian optimization, and, in a next step, find a suitable configuration for a function chain leveraging Integer Linear Programming.
Notably, modelling only requires 5-15 samples.
In contrast, Eismann et al.~\cite{2021-Eismann-Preprint-Sizeless} require significant ex-ante sampling and modelling enabling quick matching at the risk of reduced accuracy under system changes.

\subsection{Discussion}

Configuration promises matching an information system's quality goals accurately at a cost and time investment. %
To that extent, configuration methods employ three tasks sampling, modeling, and matching.
However, configuration methods can significantly differ in (i) when and how they execute each task and (ii) supported knobs, system qualities, and goals.
The fundamental design decisions that make up each configuration method are typically justified by improving accuracy or reducing the time or cost of configuration.
However, our review indicates that design decisions' negative impacts on other configuration method qualities are often hidden.
This predominantly manifests in two ways.
First, shifting negative impacts to other tasks.
For example, reporting low costs and time for matching at the expense of increased cost and time for sampling and modeling.
Second, making specific assumptions about the configured serverless system.
For example, assuming a specific workload, function composition, or FaaS platform feature to reduce efforts for sampling and modeling.

Therefore, we believe that it is beneficial to make more explicit the tradeoffs between accuracy, time, and cost that are hidden in the design decisions of current configuration methods.
In this way, tradeoffs and assumptions for a configuration method as a whole and for individual design decisions become known to researchers and practitioners, promoting (re)usability.
To this end, we propose a modular approach that allows the design decisions of different configuration methods to be combined to meet the requirements of a particular system context.
As a first step in this new direction, we propose the use of tactics, which we present in the next section.

\cfix{\vspace{-0.5em}}
\section{Tactics}\label{sec:design}\label{sec:method}\label{sec:tactics}
\cfix{\vspace{-0.5em}}
\newcommand{\Head}[1]{\vspace{0.5em}\noindent\textbf{#1}}
This section presents nine concrete tactics for engineering the quality dimensions of \method{}s.
While closely related to design patterns, we rely on the word "tactic" because they are not exclusively usable in \confsys{}s but also in semi-automated or manual \method{}s.
We group the tactics into platform-, knob-, and application-centric tactics.
We use a short-hand notation to indicate a tactic's impacts on \catdimension{}s: accuracy (\ca{}), cost (\cc{}), and time (\ct{}).  

\Head{Platform-centric Tactics}
Platform-centric tactics leverage idiosyncrasies of serverless platforms to improve a \method{}.

\tactic{T1 - Isolate Executions}
Processing multiple events concurrently in the same slot/runtime can significantly impact accuracy (\ca{}) per event by adding noise to each collected sample, thus, requiring a large number of \run{}s to filter out this noise.
Therefore, this tactic assumes that a slot only performs a single (isolated) execution at a time.
A correct assumption can reduce the number of \run{}s (\cc{}, \ct{}) without sacrificing accuracy (\ca{}); however, a false assumption runs the danger of reducing accuracy.
We observed that several FaaS platforms fulfil the assumptions of this tactic~\cite{2020-Agache-NSDI-Firecracker,2020-Kuhlenkamp-ACR-All_But_One}.

\tactic{T2 - Automate Operational Tasks}
Different \run{}s, i.e, observations, require changing the deployment and configuration of a serverless system.
This tactic uses a FaaS platform's capabilities to automate associated operational tasks.
Different FaaS platforms require a few seconds to minutes to converge to a new target deployment~\cite{kuhlenkamp-2019-ucc-opstasks}.
If a deployment change converges quickly, the lag between runs shortens, reducing the overall time for sampling (\ct).
As a downside, clients can temporarily observe inconsistent deployments in specific FaaS platforms~\cite{kuhlenkamp-2019-ucc-opstasks}.
Not accounting for this behaviour risks of mixing samples (\ca).

\tactic{T3 - Manifold Testbeds}
This tactic uses a FaaS platform's ability to scale multiple deployments independently and isolated with incoming events.
Thus, on these platforms, we can deploy \policy{}-variants in parallel and conduct multiple runs simultaneously, similar to Joyner et al.~\cite{2020-Joyner-arxiv-Ripple}.
A potential benefit is reducing the overall time of experimentation (\ct) without increasing costs (\cc) due to serverless platforms' common work-based billing model.
However, a FaaS platform's default limits, e.g., regarding the maximum number of concurrent executions, as well as cold-starts, can result in runtime bottlenecks and misleading observations impairing accuracy~(\ca).

\Head{Knob-centric Tactics}
Knob-centric tactics leverage knowledge on the relationship between a knob and its impact on quality to reduce runs while maintaining high accuracy (\ca).

\tactic{T4 - Constant Quality Function}
This tactic assumes that a system's quality does not significantly change for different values of a knob.
A \method{} can exploit this behaviour by randomly selecting a value for the knob and thus omitting runs. 
This assumption typically holds for multiple knobs exposed by FaaS platforms. 
For example, configuring \textit{functions-tags} never impacts system qualities such as latency.
Thus, omitting these knobs when considering new configuration options reduces the time (\ct{}) and cost (\cc{}).
However, validation of these assumptions is critical to avoid system quality degradation.

\tactic{T5 - Monotonic Quality Function}
This tactic assumes that values of a knob's domain have an inherent order, that quality changes with this order monotonically, and that a \method{} supports bounds in the \goal{} definition.
The method can exploit this knowledge by omitting runs after observing quality outside of a predefined bound without reducing accuracy (\ca).
While this tactic can reduce cost (\cc), it implies that runs execute sequentially, making this tactic mutually exclusive with (\manitestbeds).
For example, all elements in the domain of the \texttt{memory} knob are ordered based on their numeric value. For some applications, latency will decrease monotonically for larger memory values.
Alternatively, if a predefined quality bound states that end-to-end latency must be smaller than 1s, a method can omit runs for all smaller memory values after observing a run with a latency over 1s.

\tactic{T6 - Quality Function Type}
This tactic assumes that quality is a function of a knob's values that follows a known function type, e.g., a linear or an exponential function.
A \method{} can exploit this by not observing a \suc{} under all possible values of a knob's domain but only estimating the parameters of the known function type, which is typically possible using observations from fewer runs. 
For example, Akhtar et al.~\cite{2020-Akhtar-Infocom-COSE} assume that  execution latency as a function of \texttt{memory} follows an exponential decay function.

\tactic{T7 - Quality Function}
This tactic assumes that the function type (see \textbf{T6}), including all its parameters, are known.
Consequently, a \method{} can omit runs because the relationship between a knob and quality is known already.
In other words, this tactic (re-)uses an existing quality model entirely, omitting the tasks sampling and modelling.

\Head{Application-centric Tactics}
Application-centric tactics leverage knowledge on these decisions to reduce modelling and sampling efforts (\cc{},\ct{}) maintaining accuracy (\ca{}).

\tactic{T8 - Composition Type}
In a composition, the configuration space rapidly increases with the number of functions.
As an example, if we consider a simple sequence of three functions, the configuration space for the \texttt{memory} knob would be $10112^3$ for AWS Lambda.
However, making assumptions on the type of composition allows reducing the configuration space. 
In the context of the example, a function chain~\cite{2017-Baldini-Onward-Composition_Trilemma}, enables observing each function in isolation, followed by a suitable aggregation.
Precisely, the sum of the individual functions' execution latencies becomes an estimate of the composition's latency, thus reducing cost (\cc) and time (\ct).
This tactic finds implicit usage in~\cite{2018-Tarek-SEC-Costless} and~\cite{2020-Akhtar-Infocom-COSE}.

\tactic{T9 - Workload}
It is possible to observe significantly different qualities for \run{}s with different valid event inputs.
Consequently, maintaining high accuracy requires observing a system under different workloads, therefore, increasing the number of runs.
This tactic leverages knowledge on the impact of an event-inputs on system qualities to reduce the number of runs and benefit-cost (\cc) and time~(\ct).
Making wrong assumptions reduces the accuracy of configurations~(\ca).

\section{Sizing Middleware}\label{sec:middleware}
This section presents a novel \confsys{} (Sizer) that applies tactics from the previous section to support short lead times of serverless development.

\subsection{Requirements}\label{sec:middleware:req}
To guide the design and selection of appropriate tactics, we first define the following requirements:
A sizing request outputs a \policy{} proposing slot sizes, a request can comprise \goal{}s as bounds and preferences for the four system qualities throughput, latency, reliability, and platform consumption.
Quality preferences are a vector of relative weights that sum to one and quality bounds using comparison operators, e.g., \texttt{RLat}$\leq$900ms\footnote{total request-response latency less then 900ms}.
The Sizer shall only make minimal assumptions on a \suc{},  maintain high accuracy, defined in terms of the sum of weighted distances from an optimal policy $p^{\ast}$ (sec~\ref{sec:cat:dimensions}) using the recent proposal by Pallas et al.~\cite{2020-Pallas-BigData-Accuracy}, allowing different cloud platforms, composition types, handler implementations, and workloads.
A request shall enable short lead times by returning within minutes with a small cost per sizing.

\subsection{Tactic-based Design}\label{sec:middleware:design}
\begin{figure}[]
  \centeringß
  \includegraphics[width=0.75\linewidth]{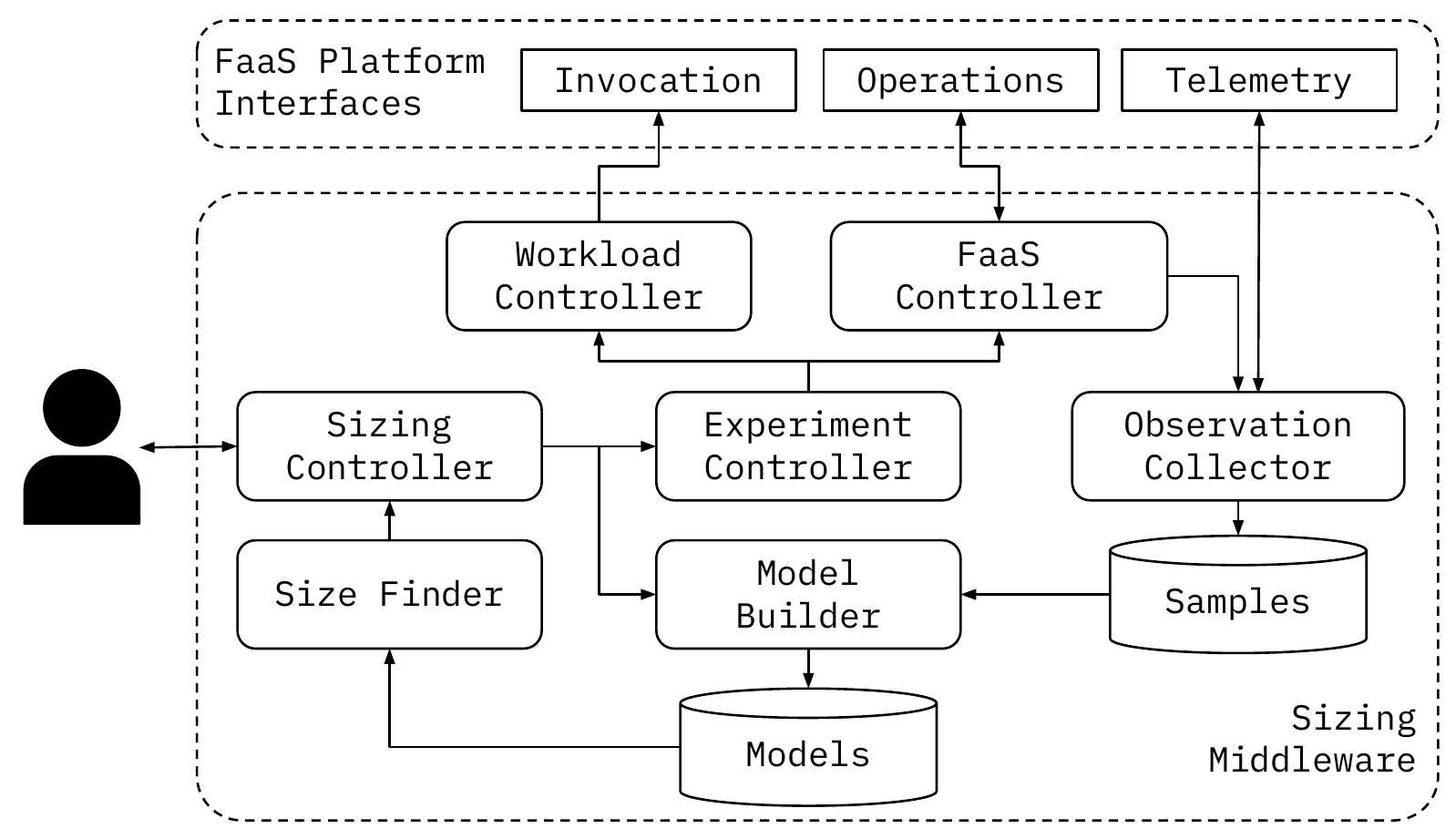}
  \caption{High-level Architecture of Sizing Middleware} 
  \label{fig:middleware-architecture}
  \vspace{-1.25em}
\end{figure}

Figure~\ref{fig:middleware-architecture} gives a high-level overview of the Sizer's architecture leverging tactics  (\textbf{T1, T2, T3, T5, T7, T8, T9}).
To utilize system information, a \textit{Sizer Client}  submits bounds, preferences, a workload model (\textbf{T9}), and a reference to a target \suc{}\footnote{For example, in the case of AWS, a references can be (i) a AWS StepFunction reference or (ii) an reference to a single Lambda function.} (\textbf{T8}) or an existing quality model (\textbf{T7}) as inputs.

As a central component, the \textit{Sizing Controller} first determines if an existing quality model can be used or if an experiment is needed to obtain a suitable quality model.
For experiments, the \textit{Experiment Controller} coordinates the generation of new \texttt{samples} leveraging a sampling strategy that prioritized time and cost,i.e., exploiting \textbf{T3} by selecting  $n_{sizes} \in \mathbb{N}$ sizes with a maximum spacing between them, instead of for example a more accurate but time-consuming approach such as Bayesian Optimization like Akthar et al.~\cite{2020-Akhtar-Infocom-COSE}.
The \textit{FaaS Controller} will enact the sampling and experimentations managed by the \textit{Experiment Controller} using the \textit{Workload Controller} which emulates a \suc{}'s clients respecting \textbf{T9}.
By further applying \textbf{T1} we can lift the need to observe different target throughputs reducing cost and time, assuming that a FaaS platform provides high elasticity~\cite{2020-Kuhlenkamp-ACR-All_But_One}.
The \textit{Workload Controller} also validates responses and invalidates deployments (\textbf{T2}). 
We primarily invalidate response of cold starts the latency introduced by cold starts is not only influenced by sizing but also other factors outside of the developers control and thus, only hurts the generated sizing model.
The \textit{Observation Controller} is than used to augment responses with added telemetry from a cloud platform, e.g., a cold start indicator.

Finally, the \textit{Sizing Controller} invokes the \textit{Model Builder} that computes quality models for each function (\textbf{T8}), and provided workload class (\textbf{T9}) and stores them in the \textit{Models} database.
This includes learning \texttt{ELat} as a function of size assuming an exponential decay function similar to Akhtar et.al.~\cite{2020-Akhtar-Infocom-COSE}.%
For compositions, observation models for each function use an aggregation function to compute the final sizing model.
An aggregation function allows to analytically determining an operation's quality only using the quality models of individual functions (\textbf{T1}).
For \texttt{cost} and \texttt{reliability}, the aggregation of compositions is straightforward, however for end-to-end latency, we need to sum the latency of sequential parts and adding the maximum latency of parallel parts of the execution flow.

Lastly, the \textit{Sizing Controller} returns the proposed \policy{} to the \textit{Sizer Client} and optionally forwards it to the \textit{FaaS Controller} that applies the \policy{} to a target deployment.

\subsection{Implementation}
We implemented the design (sec~\ref{sec:middleware:design}) in Python targeting GCF and primarily AWS to showcase applicability for FaaS platforms exposing fine-grained configuration options.
We provide a standalone executable with an integrated web-based frontend and a library for easy integration in a deployment pipeline. 
Furthermore, the Sizer's components can be used individually, such as the Model Builder and Size Finder.

We ensured that all models are file-based and storable in version-controlled systems such as Git. 
The FaaS-Controller leverages the Python-Lambda SDK for AWS, and the Observation Controller uses the AWS Cloud-Watch SDK.
We implemented custom Python logic for the Workload Controller and Composition Model and multiple Size Finder and Model Builder variants that the Sizing Controller can select.

We implemented a \texttt{Regression Sizer} that fits the exponential decay function to samples using the \texttt{scipy.optimize}\footnote{\url{https://docs.scipy.org/doc/scipy/reference/optimize.html}} package.
For the Size-Finder, we implemented a weighted sum function ($ZF$) to find Pareto-optimal solutions of the multi-objective optimization problem regarding provided the quality goals. 
For example, we assume the weights for ELat\footnote{Execution Latency} ($w_1$) and ECostt\footnote{Execution Cost} ($w_2$) and obtain for the observed requests $R$ for a size the ZF:
\begin{equation*}
 ZF(R) = \frac{1}{|R|}\sum_{r \in R} \frac{w_1 \cdot ELat_r}{\max_{a \in R} (ELat_a)}+\frac{w_2 \cdot ECost_{r}}{max_{b \in R}(ECost_b)}
\end{equation*}

For sizing compositions, the system combines quality models of individual functions to find the best \policy{} using the \texttt{scipy.optimize} package, specifically the generalized simulated annealing algorithm~\cite{1997-Xiang-Annealing}.
We published the implementation of the Sizer GitHub\footnote{\url{https://github.com/tawalaya/cat-sizer}}.

\section{Conclusion}\label{sec:conclusion}
Serverless systems require \method{}s to deal with quality-sensitive configuration options.
Results from a literature review indicate that industry and research propose isolated \method{}s that focus on accuracy but come with implicit assumptions on a system's architecture and unclear developers' time and cost investments.
We synthesized the results into nine general tactics to aid developers in the design and evaluation of \method{}s.
We do not claim completeness and invite fellow researchers to add more tactics.
For future work, we propose a modular approach that allows easily combining different tactics to meet the requirements of a particular system context.

\bibliographystyle{amsplain}
\bibliography{references}
\end{document}